\documentclass[a4paper,runningheads]{llncs}
\usepackage{makeidx}  % allows for indexgeneration
\usepackage{amssymb}
\usepackage{listings}
\usepackage{color}
\usepackage[pdftex]{graphicx}
\usepackage{array}
\usepackage{url}
\newcommand{\keywords}[1]{\par\addvspace\baselineskip
\noindent\keywordname\enspace\ignorespaces#1}
\usepackage{float}

\begin{document}
\mainmatter              % start of the contribution
\title{Improved Parallel Rabin-Karp Algorithm Using Compute Unified Device Architecture}
\titlerunning{Improved Parallel Rabin-Karp Algorithm Using CUDA}  % abbreviated title (for running head)
\author{Parth Shah\inst{1} \and Rachana Oza\inst{2}
}
\authorrunning{Parth Shah et al.}   % abbreviated author list (for running head)
\tocauthor{Parth Shah, Rachana Oza}
\institute{Chhotubhai Gopalbhai Patel Institute of Technology, Bardoli, India\\\email{parthpunita@yahoo.in}\\ \and Sarvajanik College of Engineering and Technology, Surat, India\\\email{oza.rachana7@gmail.com}
}
\maketitle              % typeset the title of the contribution
% \index{Ekeland, Ivar} % entries for the author index
% \index{Temam, Roger}  % of the whole volume
% \index{Dean, Jeffrey}

\begin{abstract}        % give a summary of your paper
String matching algorithms are among one of the most widely used algorithms in computer science. Traditional string matching algorithms are not enough for processing recent growth of data. Increasing efficiency of underlaying string matching algorithm will greatly increase the efficiency of any application. In recent years, Graphics processing units are emerged as highly parallel processor. They out perform best of the central processing units in scientific computation power. By combining recent advancement in graphics processing units with string matching algorithms will allows to speed up process of string matching. In this paper we proposed modified parallel version of Rabin-Karp algorithm using graphics processing unit. Based on that, result of CPU as well as parallel GPU implementations are compared for evaluating effect of varying number of threads, cores, file size as well as pattern size.
%                         please supply keywords within your abstract
\end{abstract}
\section{Introduction}
String matching algorithms are an important part of the string algorithms. Their task is to find all the occurrences of strings (also
called patterns) within a larger string or text. These string matching algorithms are widely used in Computational Biology,  Signal Processing, Text Retrieval, Computer Security, Text editors and many more applications \cite{APPLICATIONSINGLA}. In systems like intrusion detection or searching, string matching techniques takes more than half of the total computation time \cite{INTRUSION}. These string matching algorithms can be divided mainly into two sub category: single string matching algorithms and multiple string matching algorithms. In this paper we focus on mainly multiple string matching algorithms. In multiple string matching algorithm if we want to check only if pattern in text or not then Boyer-Moore performs best, but if we want to find all occurrence of pattern in text then Rabin-Karp provides most optimal results \cite{COMPARISION}.
\let\thefootnote\relax\footnotetext{The final publication is available at Springer via \url{https://doi.org/10.1007/978-3-319-63645-0_26}}

As the information technology spreads fast, data is growing exponentially. This requires to scale up processing power which is not possible using CPU. CPUs were basically designed for general purpose work. So, it has higher operating frequency and lower number of cores. On the other end Graphics processing units were originally developed for preforming highly parallel operations like graphics rendering. For performing this large amount of computation GPUs have higher number of processing elements (ALUs) and lower operating frequency in compared to normal CPUs. If we able to use these highly parallel processing elements for our pattern matching task, we can greatly improve performance of string matching algorithm.

  In this paper, we have implemented modified parallel Rabin-Karp string matching algorithm over GPU where most of computation task is performed on GPU instead of CPU. Previous work by Nayomi at el. uses RB-Marcher technique to find hash while our approach uses Rabin Hashing as hash mechanism. The main contribution of this paper is improved parallel version of Rabin Karp algorithm that utilize inherent parallelism of Rabin Karp algorithm using GPU. In this paper, we have also evaluated performance of our modified approach over different parameters like pattern length, number of threads, filesize as well as number of cores.

Section 2 of this paper describes Related Work in detail. Algorithms and Implementation, Results and Discussion, Conclusion are discussed in
Sections 3, 4 and 5 respectively.

\section{Related work}

Accelerating string matching algorithms are one of the major concern in areas where string matching algorithm used heavily. In the era of big data this will greatly speed up the processing task. With the introduction of General Purpose Graphics Processing Units (GPGPUs) we can achieve highly parallel processing capabilities \cite{CUDA:GPGPU}. Using this parallelism we can speedup the task of string matching in efficient manner.

For the task of string matching various algorithms were developed in past like Brute-Force algorithm, Boyer-Moore algorithm, Knuth-Morris-Pratt algorithm, Rabin-Karp algorithm, etc. Blandón and Lombardo \cite{COMPARISION} had compared all these algorithm and found that Boyer-Moore \cite{Boyer:1977:FSS:359842.359859} performs best when there is no pattern matched in text. Knuth-Morris-Pratt \cite{KMP:doi:10.1137/0206024} algorithm heavily depends on result previous phase's calculation which makes it unsuitable to parallel implementation. That leads us to Rabin-Karp \cite{RABINKARP:5390135} algorithm which is inherent parallel in its design. In Rabin-Karp algorithm, calculation of hash for every substring does not depends on any other information then substring. So we can easily parallelize this operation over GPU which helps us to improve performance of Rabin-Karp algorithm.

For parallel implementation of Rabin-Karp algorithm, Nayomi at el. \cite{RABINONCUDA:7069589} proposed Rabin-Karp algorithm which uses RB-Matcher \cite{chillar2008rb} for generating hash value. But due to involvement of modular arithmetic in RB-Matcher performance improvement is limited. In our work, we have presented GPU version of Rabin-Karp algorithm that uses Rabin Hashing as hash function. Our work not only just propose modified approach, but also proves that using experimental results on real data.

\section{Algorithm and Implementation}

In this section, we discuss about modified Rabin-Karp algorithm and our parallel CUDA implementation details.
\subsection{Rabin-Karp algorithm}

Rabin Karp is a string searching algorithm, which supports both single and multiple pattern matching capabilities. It is widely used in applications like plagiarism detection and DNA sequence matching.

In our proposed approach we have used Rabin hashing method to find out pattern strings in a text. In order to compute hash we apply left shift to ASCII value of character and add it to previously calculated hash. We apply this process repeatedly for all character of string. To find out the matching pattern we will compare hash of pattern and hash of each substring of text. If hash of both substring and pattern match then only we will compare string to find match.

\medskip
\noindent
{\it Pseudo code for sequential implementation}
\begin{verbatim}
function Rabin_Karp_CPU(result, T, n, P, m)
{
    hx=0,hy= 0;
    for (i = 0;i<m;++i)
    {
        hx = ((hx << 1) + P[i]);
    }
    for (x = 0;x < n - m;x++)
    {
        for (hy = i = 0;i <m; ++i)
        {
            hy = ((hy << 1) + T[i + x]);
        }
        /*  Compare hash and match pattern  */
        if (hx == hy	&&	compare(P, T + x, m) == 0)
        {
            print("Match at:", x);
            result[x] = true;
        }
    }
    return result;
}
\end{verbatim}

In sequential implementation pattern matching function is called for each substring sequentially. First we read text T and pattern P from file. Text T of length n and pattern P of length m is passed as argument to matching function. Function first calculate hash of pattern hx and store it in memory. After that for each substring of length m in text T, hash is calculated. If hash of both pattern and substring matches then only we will compare string. If string is matched we will store its index in result array. And result is returned with containing matching element.

\medskip
\noindent
{\it Pseudo code for parallel GPU implementation}
\begin{verbatim}
function Rabin_Karp_GPU(T, P, n, m, hx, result)
{
    blockId = blockIdx.x + blockIdx.y * gridDim.x +
              gridDim.x * gridDim.y * blockIdx.z;
    x = blockId * blockDim.x + threadIdx.x;
    if (x <= n - m)
    {
        hy = 0;
        for (i = 0;i <m; ++i)
        {
            hy = ((hy << 1) + T[i + x]);
        }
        /*  Compare hash and match pattern  */
        if (hx == hy	&&	compare(P, T + x, m) == 0)
        {
            print("Match at:", x);
            result[x] = true;
        }
    }
}
\end{verbatim}

In parallel version, we first load text T and pattern P from file to global memory. Once pattern is loaded in memory we calculate hash hp. Once hp is calculated we pass it as argument to string matching function. At the same time T and P are transferred to GPU memory so that we can access it using GPU's processing elements. String matching function will execute for each substring in parallel on GPU. Processing elements on GPU can be access using grids. Each grid have multiple blocks where each block can execute upto 1024 threads in parallel. We will calculate offset of string index by combining id of thread, block and grid which is stored as x in our pseudo code. Based on offset we calculate hash hy of each substring on individual GPU thread. If hash of both substring and pattern matches then only string is compared. If string matches offset is stored in result and result is transferred back to CPU memory.

\subsection{CUDA Architecture}
Compute Unified Device Architecture (CUDA) is developed as a parallel computing and application architecture by NVIDIA corporation. It enables user to harness the computing capacity of GPUs. It allows us to send our C / C++ program directly to GPU for execution. Tools for interacting with GPUs having CUDA architecture is available as CUDA Toolkit released by NVIDIA. In this article, CUDA toolkit and CUDA enabled GPUs are used for implementation of parallel GPU version of Rabin-Karp algorithm.
\vspace{-10px}
\begin{figure}
\includegraphics[width=12cm,height=6cm]{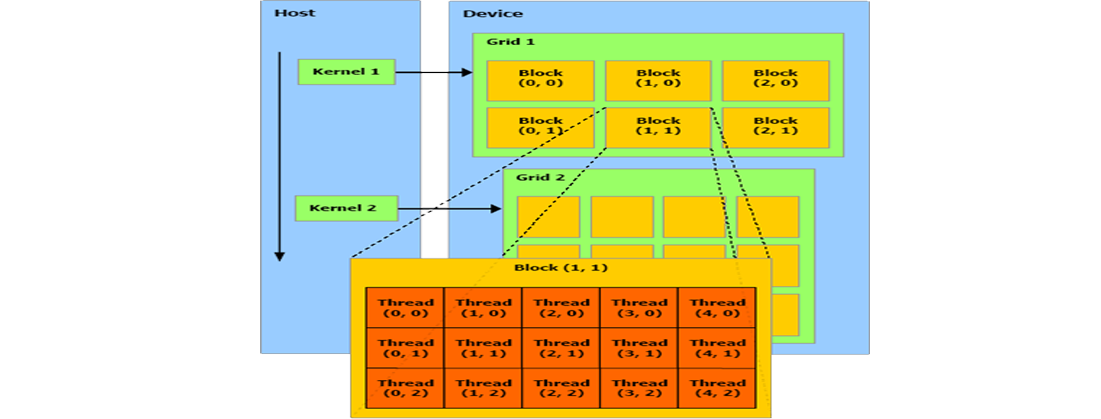}
\caption{Architecture of CUDA \cite{NVIDIAINTROCUDA:WILL}}
\end{figure}
Some terminologies used in NVIDIA CUDA programming environment is as follows:\\
      \textbf{Host:} The CPU will be referred to as a host.\\
      \textbf{Device:} An individual GPU will be referred to as a device.\\
      \textbf{Kernel:} Kernel is a function that will be executed in individual block.\\
      \textbf{Thread:} Each thread is an execution of a given kernel with a specific index. Each thread uses its index to access elements in array.\\
      \textbf{Block:} A group of threads that are executed together and form the unit of resource assignment is called block.\\
      \textbf{Grid:} Grid is a group of independently executable blocks. \\
Our main algorithm for parallel implementation is used as a kernel function. CUDA architectures support maximum of 1024 threads per block. Size of block and grids are selected such that multiplication of number of thread, block and grid will match the total characters in text. This will helps us to effectively optimize all cores of GPU.

\subsection{Hardware and Datasets}

All the experiment in this paper carried out using Intel Core i7 processor with 4 cores. For GPU implementation we have systems with two different number of cores as increasing number of cores increase total available processing elements. They are as follows: 1) NVIDIA GT920 with 386 CUDA cores and 4GB Memory and 2) NVIDIA GForce GT 630M with 96 CUDA cores and 2GB Memory.

For evaluation of algorithm we have used Random DNA sequence generator to generate dna sequence dataset. We have used dataset of different sizes like 2MB, 10MB, 20MB, 40MB, 80MB, 160MB, 360MB, 640MB and 1GB.

\section{Result and Discussion}
For evaluating performance of our proposed parallel algorithm over serial version we have used speed up ratio as our performance measure. Speed up ratio S is calculated as,
\begin{equation}\label{eq1}
  S = \frac{T_{GPU}}{T_{CPU}}
\end{equation}
where $T_{GPU}$ and $T_{CPU}$ are execution time for GPU and CPU respectively.

First we have compared our parallel algorithm over sequential algorithm. For that we have executed our modified algorithm with different number of threads in parallel execution.
\begin{table}[h]
\vspace{-5px}
\caption{Performance comparison for different thread size}\label{table2}
\centering{}%
\begin{tabular}{>{\centering}m{3cm}>{\centering}m{3cm}>{\centering}m{3cm}>{\centering}m{3cm}}
\hline
Number of Threads Per Block  & Sequential Implementation  & Parallel GPU implementation  & Speed Up Ratio \tabularnewline
\hline
32 & 6937 & 4718.750 & 1.470093\tabularnewline
64  & 6535 & 2708.375 & 2.412886\tabularnewline
128 & 6839 & 1715.790 & 3.985919\tabularnewline
256 & 6360 & 1220.096 & 5.212705\tabularnewline
512 & 6047 & 971.2466 & 6.226019\tabularnewline
1024 & 6125 & 847.1143 & 7.230430\tabularnewline
\hline
\end{tabular}
\vspace{-10px}
\end{table}

As we can see in Table \ref{table2} as number of threads increases total execution time required for parallel implementation will reduce proportionally.

To check how our algorithm scales when we increase number of cores we have implemented our parallel algorithm on GPUs with 96 and 384 cores. Result of execution on both GPU keeping constant pattern size of 7 over 320MB filesize is given in Table \ref{table4}.

\begin{figure}[h]
\vspace{-10px}
\centering{}\includegraphics[width=12cm,height=4.2cm]{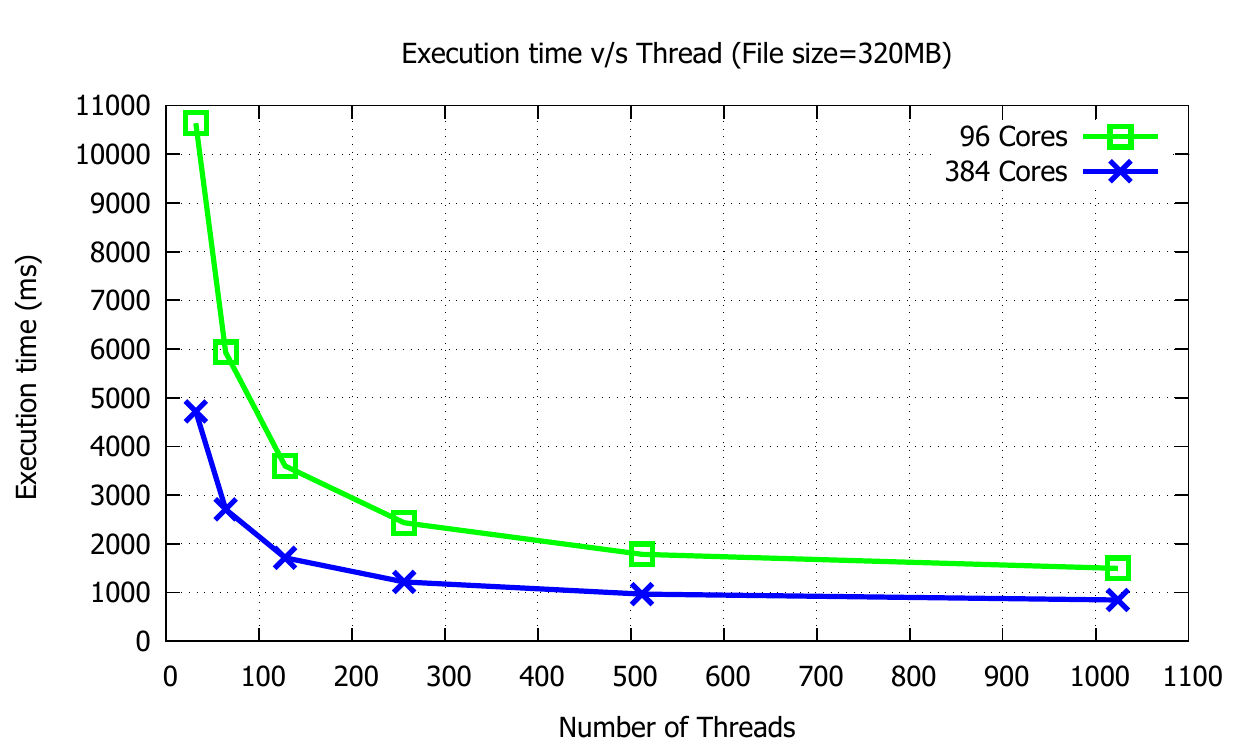}\caption{Performance comparison for different number of cores}\label{fig94_384}
\end{figure}

\begin{table}[h]
\vspace{-10px}
\caption{Performance comparison for different number of cores}
\label{table4}
\centering{}%
\begin{tabular}{>{\centering}m{3cm}>{\centering}m{3.5cm}>{\centering}m{3cm}}
\hline
Number of Threads Per Block  & NVIDIA GForce GT630M (96 cores) (ms) & NVIDIA GForce 920 (384 cores) (ms) \tabularnewline
\hline
32 & 10638.17  & 4718.75\tabularnewline
64 & 5931.667  & 2708.375 \tabularnewline
128 & 3600.603  & 1715.79 \tabularnewline
256 & 2434.065  & 1220.096 \tabularnewline
512 & 1787.352  & 971.2466 \tabularnewline
1024 & 1495.54  & 847.1143 \tabularnewline
\hline
\end{tabular}
\vspace{-10px}
\end{table}
As we can see from Fig. \ref{fig94_384}, increases in speed up ratio is directly proportional to number of cores in GPU. As we increase number of cores total time require for execution will decrease proportionally.

In order to check effect of pattern size on execution time, we executed both sequential and parallel version using different pattern size.  As we can see from Table \ref{ref1}, increasing input pattern size increases speed up ratio due to fact that GPUs can easily handle computation heavy task with their higher number of cores compared to CPU.

\begin{table}[h]
\caption{Performance comparison for different pattern length}\label{ref1}
\centering{}%
\begin{tabular}{>{\centering}p{3cm}>{\centering}p{3cm}>{\centering}p{3cm}>{\centering}p{3cm}}
\hline
Pattern Length  & Intel Core i7 (4 Cores) (ms)  & NVIDIA GForce 920 (384 cores) (ms)  & Speed Up Ratio \tabularnewline
\hline
25 & 140 & 7.51616 & 18.62653\tabularnewline
50 & 328  & 13.82605 & 23.72334 \tabularnewline
100 & 578  & 26.37414  & 21.9154 \tabularnewline
200 & 1187 & 51.32698 & 23.12624 \tabularnewline
800 & 4578 & 201.1771 & 22.75607\tabularnewline
\hline
\end{tabular}
\end{table}

\begin{figure}[H]
\vspace{-25px}
\centering{}\includegraphics[width=12cm,height=4.5cm]{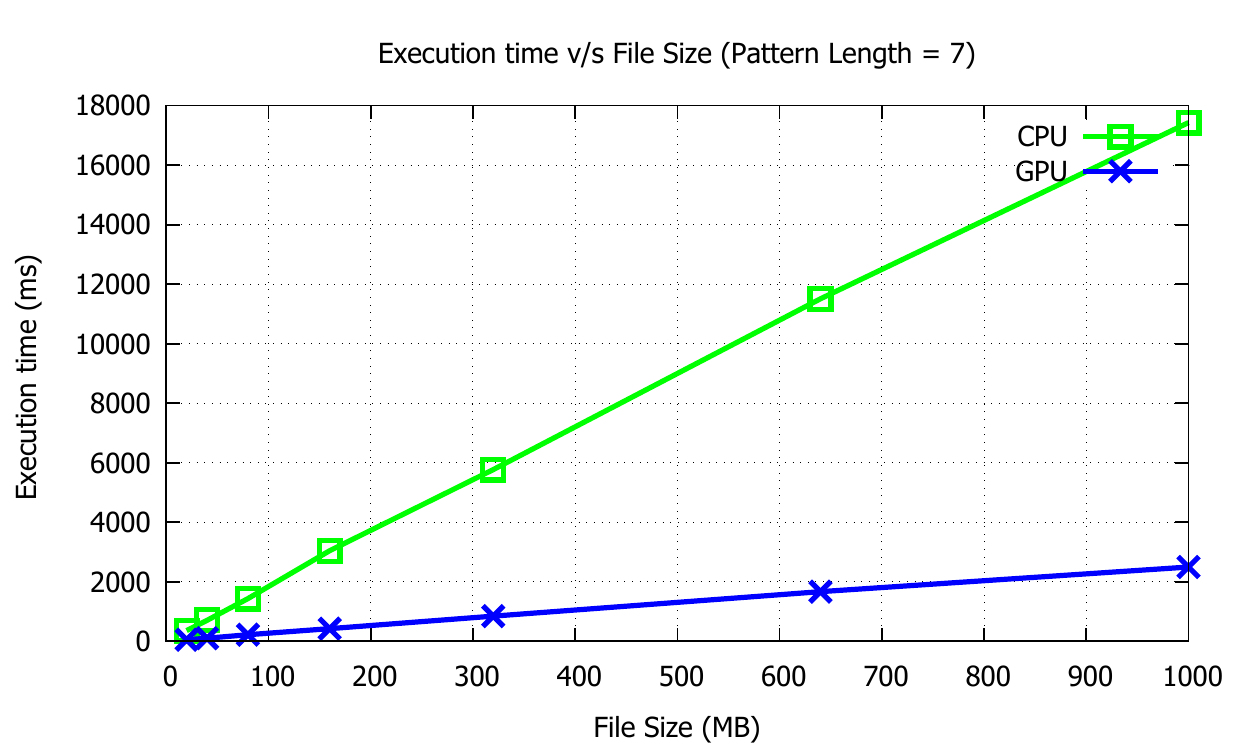}\caption{Performance comparison for different file size}\label{fig3}
\vspace{-20px}
\end{figure}

As our main goal is to process and match pattern in real life situations where we encounter different file sizes. So to check performance, we have evaluated performance of our GPU implementation for different filesize in Fig. \ref{fig3}. As we can see, GPU version of parallel algorithm performs better when file size is larger as increasing file size dramatically increase computation which can easily handled by GPU.

\section{Conclusion}
Comparing implementation of serial and CUDA version of Rabin-Karp string matching algorithm, we can get upto 23x speed-up in CUDA version over serial version. Using Rabin hashing reduces total computation required for generating hash which directly reflect in execution speed. From our experiments we can conclude that maximum speedup is archived when file size is minimum and as file size increase speedup decreases. Similarly by increasing number of cores and execution threads in GPU we can get maximum speedup in task of string matching as it increase total number of tasks that can be executed concurrently.


\bibliographystyle{splncs}
\bibliography{reference_final}
\end{document}